\documentclass[11pt,onecolumn,journal]{IEEEtran}
\ifCLASSINFOpdf
\else
\fi
\usepackage[cmex10]{amsmath}

\usepackage{microtype}
\usepackage{graphicx}
\usepackage{subfigure}
\usepackage{graphics} 

\usepackage{amsthm}
\usepackage{amssymb} 
\usepackage{xspace}
\newtheorem{theorem}{Theorem}

\newtheorem{lemma}{Lemma}
\newtheorem{remark}{Remark}

\usepackage{color, colortbl}

\newtheorem{definition}{Definition}
\newcommand{\F}{\ensuremath{\mathbb F}}
\newcommand{\Fc}{\ensuremath{\mathcal F}}
\definecolor{Gray}{gray}{0.9}
\newcommand{\V}{\ensuremath{\mathbb V}}
\newcommand{\Uu}{\ensuremath{\mathbb U}}
\usepackage{wrapfig}
\usepackage{lipsum}
\newcommand{\wt}{worker threshold\xspace}
\newcommand{\Wt}{Worker threshold\xspace}
\newcommand{\WT}{Worker Threshold\xspace}
\newcommand{\hs}{homogeneous scheme\xspace}
\newcommand{\ns}{non-homogeneous scheme\xspace}
\newcommand{\workers}{w}
\newcommand{\K}{k}
\newcommand{\byz}{b}
\newcommand{\ls}{e}
\newcommand{\mls}{e}
\newcommand{\lin}{l}
\newcommand{\parti}{T}
\newcommand{\partisize}{t}
\newcommand{\matpartsize}{t}
\newcommand{\cics}{intersecting input curves scheme\xspace}
\newcommand{\inob}{input oblivious\xspace}
\newcommand{\repeatedExponent}{{\langle s+1 \rangle}}
\newcommand{\localRecovery}{local recovery\xspace}
\usepackage{pgfplots}
\pgfplotsset{compat=1.9}

\begin{document}

\title{A locality-based approach for coded computation}

\author{Michael~Rudow, K.V.~Rashmi, and Venkatesan~Guruswami
\thanks{M. Rudow, K.V. Rashmi, and V. Guruswami are with the Computer Science Department, Carnegie Mellon University}
}

\maketitle

\thispagestyle{empty}
\pagestyle{empty}

\begin{abstract}
Modern distributed computation infrastructures are often plagued by unavailabilities such as failing or slow servers. 
These unavailabilities adversely affect the tail latency of computation in distributed infrastructures. 
While replicating computation is a simple approach to provide resilience, it entails significant resource overhead. 
Coded computation has emerged as a resource-efficient alternative, wherein multiple units of data are encoded to create parity units and the function to be computed is applied to each of these units on distinct servers. 
If some of the function outputs are unavailable, a decoder can use the available ones to decode the unavailable ones. 
Existing coded computation approaches are resource efficient only for simple variants of linear functions such as multilinear, with even the class of low degree polynomials necessitating the same multiplicative overhead as replication for practically relevant straggler tolerance.

In this paper, we present a new approach to model coded computation via the lens of locality of codes. We introduce a generalized notion of locality, denoted \textit{computational locality}, building upon the locality of an appropriately defined code. We then show an equivalence between computational locality and the required number of workers for coded computation and leverage results from the well-studied locality of codes to design coded computation schemes. 
Specifically, we show that recent results on coded computation of multivariate polynomials can be derived using \localRecovery schemes for Reed-Muller codes. 
We then present coded computation schemes for multivariate polynomials that adaptively exploit locality properties of input data\textemdash an inadmissible technique under existing frameworks. These schemes require fewer workers than the lower bound under existing coded computation frameworks, \textit{showing that the existing multiplicative overhead on the number of servers is not fundamental for coded computation of nonlinear functions.}  

\end{abstract}


\section{Introduction}

Large scale computations involving smaller modular pieces are widely prevalent. The quintessential example of this is computationally intensive distributed machine learning applications. Distributed computations suffer from unavailabilities such as ``straggling'' servers (i.e. servers that fail or are slow to respond) \cite{dean2013tail}. These unavailabilities result in an increase in the tail latency for the distributed computations.

One natural approach to address server unavailabilities is to add redundancy, such as replicating computations. However, using replication requires a high resource overhead, indicating the need for more efficient techniques. In the setting of distributed storage, it is well known that the technique of erasure coding provides robustness to failed nodes with minimal storage overhead \cite{weatherspoon2002erasure}. This efficient performance has motivated the strategy of similarly using erasure coding to provide robustness to failing servers (or workers) for distributed computation; this approach is termed ``coded computation'' \cite{lee2018speeding}. Specifically, coded computation involves querying workers with encoded inputs so that the missing outputs from unreliable workers are recoverable from the received worker outputs.

Coded computation was initially introduced in~\cite{lee2018speeding} for matrix multiplication operations. Several subsequent works have presented various coded computation approaches applicable to linear computations \cite{li2016unified,reisizadeh2019coded,mallick2019rateless,lin2019train,dutta2017coded,dutta2016short,ferdinand2018hierarchical,dutta2018unified}, bilinear matrix-matrix multiplication computations \cite{lee2017high,yu2017polynomial,dutta2019optimal,wang2018sparse,baharav2018straggler,kiani2018exploitation,jeong2018locally,narra2019slack,yu2018straggler,park2018hierarchical,Jia2019cross,yu2020entangled}, and multilinear matrix multiplication of more than two matrices \cite{dutta2019optimal}. In ~\cite{yu2018lagrange}, Yu et al. designed a coded-computation scheme for multivariate polynomial functions. Under the model considered in \cite{yu2018lagrange}, the authors also proved a lower bound for coded computation of multivariate polynomials of degree at least $s+1$ tolerating up to $s$ stragglers, necessitating a \textit{factor $s$ overhead on the required number of workers}. 
Thus, using existing approaches, providing resilience to even $1$ (or $2$) stragglers for degree $2$ or $3$ multivariate polynomials would require at least $100\%$ (or $200\%$) overhead on the number of workers.

Motivated by large resource overhead for coded computation of nonlinear functions under existing approaches, we consider a \textit{new approach to model coded computation using the concept of locality of codes}. To do so, we define a new notion of locality for computation, denoted \textit{computational locality}, via locality properties of an appropriately defined code. The topic of the locality of codes has a rich literature~\cite{locally-decodable-codes,gopalan2012locality,papailiopoulos2014locally,
tamo2014family,prakash2014codes,ramakrishnan2018taking}. 
For instance, there are well known ``\localRecovery schemes''\textemdash procedures to recover one erased code symbol by querying a few other code symbols, for several classes of codes such as Reed-Muller codes. 
However, conventional study of locality of codes typically differs from the coded computation setting in two main respects: (1) The coded computation setting admits replicated queries, whereas such replicated queries are not available under typical locality regimes. (2) The coded computation setting handles arbitrarily many inputs whereas traditional study of locality usually considers one (or sometimes two) inputs. The main contributions of this paper are twofold:

First, we propose a new model for coded computation via the lens of locality of codes. 
We demonstrate that the proposed notion of computational locality of functions is equivalent to the required number of workers for coded computation of that function. We then show how to use the proposed locality-based model to exploit \localRecovery schemes of codes to design coded computation schemes. The locality-based model, thus, enables the domain of coded computation to exploit the well-studied notion of locality of codes. Specifically, we show that the \localRecovery schemes for Reed-Muller codes can be generalized to yield a coded computation scheme for multivariate polynomials. The scheme so obtained reproduces the best known coded computation scheme for multivariate polynomial functions \cite{yu2019lagrange} from an alternative viewpoint, that of the locality of Reed-Muller codes. We also show that proposed locality-based model can be used to interpret existing coded computation schemes in other settings as well.

Second, in addition to providing a unified perspective, the locality-based approach to modeling coded computation circumvents existing lower bounds on the required number of workers. 
By exploiting the proposed locality-based model, we design a coded computation scheme for multivariate polynomial functions using fewer workers than the minimum required number under previously studied models. Specifically, the proposed scheme provides robustness to $s$ stragglers for any degree $(s+1)$ multivariate polynomial using less than a factor $s$ of extra servers. This establishes that \textit{a factor $s$ overhead in the number of workers is not a fundamental requirement for coded computation of non-linear functions.} This opens up the potential for resource efficient coded computation schemes for non-linear functions. We design the coded computation scheme via a two-step process: First, we design a scheme for homogeneous multivariate polynomial functions which adaptively exploits linear dependencies of the input data to reduce the number of workers. Second, we extend the scheme to apply to non-homogeneous multivariate polynomial functions via the technique of homogenizing polynomials. We then discuss regimes where the locality-based approach to modeling coded computation admits significant reduction in the required number of workers. Finally, we consider the general class of coded computation techniques enabled by the locality-based model of coded computation.

\section{Background and Related Works}
\label{sec:background}
\begin{figure}
\centering

\includegraphics[scale=.48]{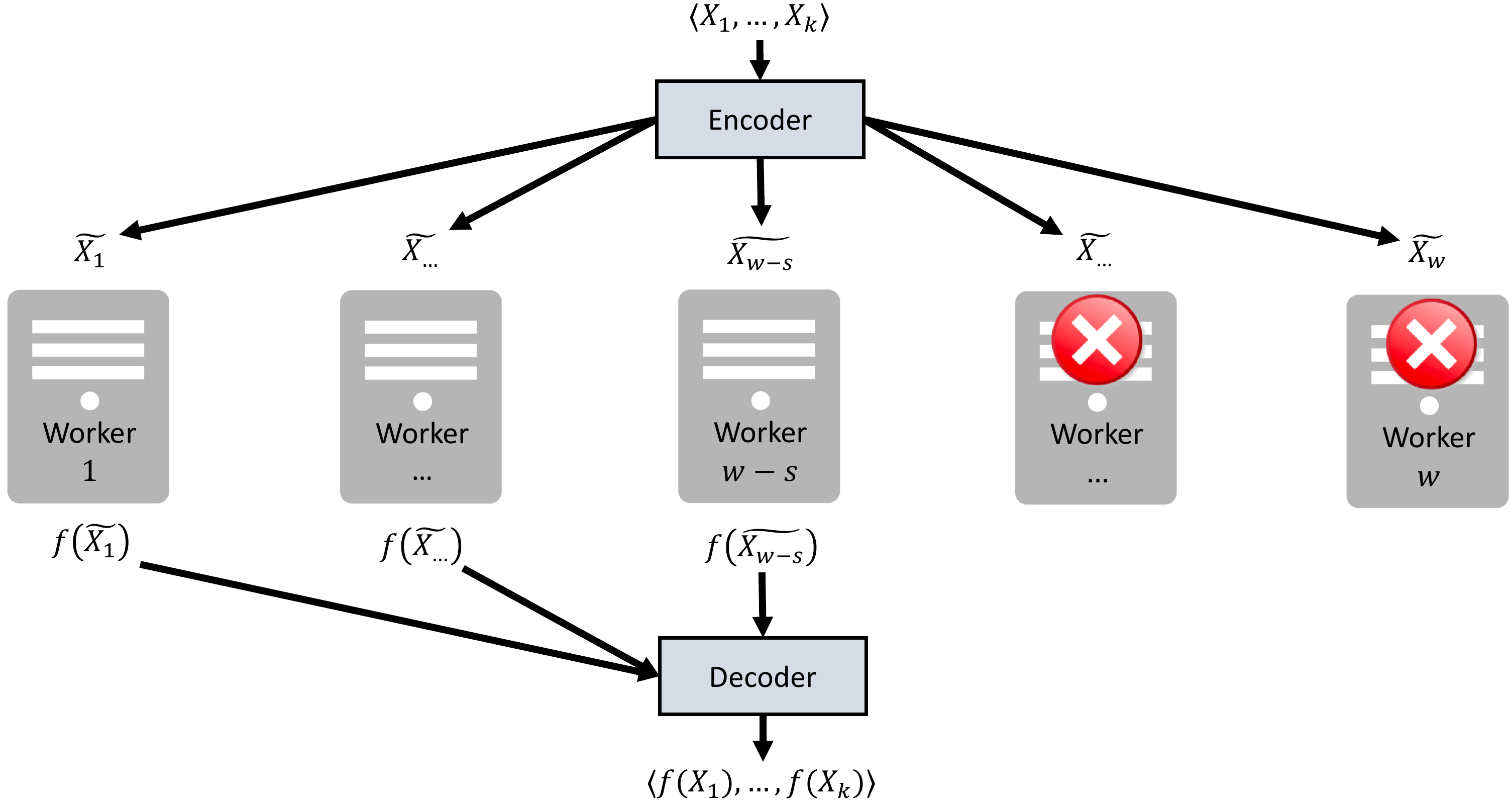}
\caption{Overview of the coded computation model. A master uses $w$ workers to compute the value of the function $f$ at $\K$ evaluation points subject to at most $s$ straggling workers.}

\label{fig:setting}

\end{figure}

We start by describing the coded computation model studied in this work which is based on the model considered in \cite{yu2019lagrange}.

\textbf{Setting:} The setting is shown in Figure~\ref{fig:setting} and consists of one master and many workers. The master's objective is to compute the value of a function $f$ at $\K$ arbitrary input points $\langle X_1,\ldots,X_\K \rangle$ using the minimum necessary number $\workers$ of workers. Moreover, the function $f$ can be any function in a set of functions $\Fc$\textemdash henceforth denoted as a ``function class''\textemdash and up to $s$ arbitrary workers may straggle (fail).\footnote{The model is later extended to include up to $\byz$ byzantine (adversarial) workers in addition to the at most $s$ straggling workers.} To do so, the master ``queries'' each worker $i \in \{1,\ldots,\workers\}$ with the evaluation point $\widetilde{X_i}$, and the worker computes $f(\widetilde{X_i})$. At most $s$ arbitrary workers straggle and return nothing to the master; the remaining $\workers-s$ workers return their computed values to the master. The master must use the received subset of $\{f(\widetilde{X_1}),\ldots,f(\widetilde{X_\workers})\}$ to decode $\langle f(X_1),\ldots,f(X_\K)\rangle$ regardless of which worker outputs are missing. The threshold of the minimum number of workers which is sufficient to do so is defined as follows:

\begin{definition}[\Wt]
The \textbf{$(\Fc,\K,s)$-\wt} is the minimum number $\workers_{\Fc,\K,s}$ of workers for which coded computation for any function $f$ in a function class $\Fc$ at any $\K$ input points tolerating up to $s$ straggling workers is possible.
\end{definition}

\subsection{Related Works}
The domain of coded computation has received considerable attention in the recent past. We now briefly overview some of the relevant topics.

Most works have considered coded computation for functions within the domain of Cartesian products of multivariate polynomial functions. For example, many works have considered linear settings such as matrix-vector multiplication \cite{lee2018speeding,reisizadeh2019coded,mallick2019rateless,lin2019train,li2016unified,ferdinand2018hierarchical,dutta2018unified}, convolution~\cite{dutta2017coded}, linear transforms~\cite{dutta2016short}, etc. Several works have considered coded computation for bilinear matrix-matrix multiplication~\cite{lee2017high,yu2017polynomial,dutta2019optimal,wang2018sparse,baharav2018straggler,kiani2018exploitation,jeong2018locally,narra2019slack,yu2018straggler,park2018hierarchical,Jia2019cross,yu2020entangled}, and, in \cite{dutta2019optimal}, the authors designed a coded computation scheme to compute the matrix product of any number of matrices. 
In~\cite{yu2019lagrange}, the authors proposed a coded computation scheme for the Cartesian product of finitely many multivariate polynomial functions. Due to its generality, this scheme is portable; for example, it is is used in \cite{yang2019timely} for a setting involving multiple rounds of computation. Understanding coded computation for the large class of polynomial functions is a significant milestone in moving towards the ultimate goal of general functions.

Robust distributed computation for machine learning has received substantial interest. Most existing work focuses on the training phase, designing coded approaches for gradient computations during gradient descent. As the gradient is the summation of ``partial gradients,'' several works compute the partial gradients in a coded manner~\cite{tandon2017gradient,halbawi2018improving,li2018near,chen2018draco,charles2018gradient,ye2018communication}. Other works designed robust schemes to compute an approximate value of the gradient~\cite{raviv2018gradient,wang2019erasurehead,charles2017approximate,charles2018gradient,karakus2017straggler}.

In~\cite{kosaian2018learning}, Kosaian et al. introduced coded computation for neural network inference. Providing resilience to inference tasks via coded computation requires supporting non-linear functions, as neural networks are highly non-linear. The authors propose a learning-based approach to overcome the challenge of non-linearity by using machine learning to learn an encoder and decoder for any given computation. Other works~\cite{kosaian2019parity,narra2019collage} have further explored this learning-based approach for coded computation of inference tasks.

The learning-based approach approximately reconstructs unavailable function outputs. While approximation is suitable for ML inference tasks, it may not be suitable for others. Even ML tasks can benefit from exact reconstruction when it is achieved efficiently. This motivates pursuing mathematical constructs for coded computation for non-linear functions, as we do in this work.

\subsection{Coded Computation of Multivariate Polynomials}
We now describe the coded computation setting studied in \cite{yu2019lagrange} in more detail, as it is closely related to the results presented in this work. This previously studied coded computation model is a special case of the setting described in the beginning of Section~\ref{sec:background} with the following restrictions: (1) Coded computation is restricted to an arbitrary multivariate polynomial functions $f: \mathbb{V} \rightarrow \mathbb{U}$ for vector spaces $\mathbb{V}$ and $\mathbb{U}$ over a field $\F$. (2) The master queries each worker at an evaluation point which is a linear combination of the input points $\langle X_1,\ldots,X_\K\rangle$. Specifically, for each $i \in \{1,\ldots,\workers\}$, worker $i$ computes the value of $f(\cdot)$ at the evaluation point $\widetilde{X_i} = \sum_{j=1}^\K \alpha_{i,j} X_j$ where $\alpha_{i,j} \in \F$. Moreover, the $\alpha_{i,j}$ are fixed apriori and are the same for any input points $\langle X_1,\ldots,X_\K\rangle$. (3) The master decodes $\langle f(X_1),\ldots,f(X_\K)\rangle$ using a linear combination of the received worker outputs. As the same
linear encoding and decoding schemes are used for any $\K$ input points, we denote this setting as \textit{\inob}.

\textbf{Lagrange Coded Computing (LCC) scheme \cite{yu2019lagrange}}: The scheme computes the value of a multivariate polynomial function, $f:\mathbb{V} \rightarrow \mathbb{U}$, at input points $\langle X_1,\ldots,X_\K\rangle$ while tolerating up to $s$ straggling workers. When $deg(f) \le (s+1)$, the LCC scheme computes the univariate polynomial $f(p(z))$, where $p(z)$ is a univariate polynomial of degree at most $\K-1$ passing through the $\K$ input points. Specifically, for $1 \le i \le \K$ and distinct $\beta_1,\ldots,\beta_{\K} \in \F$, $p(\beta_i) = X_i$. The scheme computes $f(p(z))$ by using each of the $\workers=deg(f(p(z)))+s+1$ workers to compute the value of $f(p(z))$ at a distinct point. It then uses polynomial interpolation to recover $f(p(z))$ while tolerating any $s$ straggling workers. Finally, the scheme decodes each $f(X_i) = f(p(\beta_i))$.  When $(s+1)<deg(f)$, the LCC scheme uses $(s+1)$-wise replication with $\workers = \K(s+1)$ workers and direct decoding. The authors prove a lower bound for this \inob setting, showing that any scheme requires at least $min(\K(s+1), (\K-1)deg(f)+s+1)$ workers.

\section{Locality-Based Model for Coded Computation}
\label{sec:prelim}

In this section, we propose a new locality-based model for coded computation; this model provides a unified viewpoint for coded computation and enables reduced \wt in the domain of multivariate polynomial computations compared to existing models. 
\subsection{Comparison to Existing Model}
The two key aspects of the locality-based model of coded computation considered in this work are: 
First, for any $\K$ input points, the master may query each worker at an arbitrary evaluation point. 
Second, the decoding function can be an arbitrary function of the received worker outputs. 
This setting is a generalization of the model considered in \cite{yu2019lagrange}; in the model used in \cite{yu2019lagrange}, the coded computation scheme is required to use the same linear encoding and decoding scheme for every $\K$ input points and is therefore denoted as \inob. The flexibility to adapt to specific input points is fundamental to exploit locality properties of the input points, as we will see later.

In this setting, the computational complexity of the master (for encoding and decoding) can be higher than in the \inob setting, while the computational complexity of the workers remains the same in both settings. Given that there is only one master while there are numerous workers, we argue that the reduction in the number of workers needed justifies the increase in computational complexity of the master. In certain applications, the extra work of the master may be amortized over multiple rounds of coded computation. For example, in gradient descent for machine learning applications, the gradient is computed at the same points over multiple rounds. Thus, the master might identify locality properties of the input points (as will be described in Section~\ref{sec:Local}) in the first iteration and reuse these locality properties in later rounds at no additional cost.

\subsection{Preliminaries}

We now provide a few relevant definitions from coding theory. In these definition and later in this work, $[i]$ denotes $\{1,\ldots,i\}$. Moreover, when considering the Cartesian product of multivariate polynomial functions $f(X) = (f_1(X),\ldots,f_{m}(X))$, the convention of $deg(f)=max_{i \in [m]} deg(f_i)$ is used.

\begin{definition}[Distance]
Codewords $c_1$ and $c_2$ of a code $C$ have \textbf{distance} $\Delta(c_1,c_2)$ of the number of symbols in which $c_1$ and $c_2$ differ. 
\end{definition}

\begin{definition}[Hamming ball]
For a codeword $c$ of a code $C$, the \textbf{Hamming ball} $B(c,r)=\{c' \in C \mid \Delta(c,c') \le r\}$.  
\end{definition}

\begin{definition}[Punctured code]
For a code $C \subseteq \Uu^n$ and a subset $I =\{i_1,\ldots,i_l\} \subseteq [n]$, the \textbf{punctured} code $C_I=\{c_I = (c_{i_1},\ldots,c_{i_l}) \mid c=(c_1,\ldots,c_n) \in C\}$ is the code formed by restricting the codewords of $C$ to the symbols corresponding to $I$.
\end{definition}

\subsection{Computational Locality}

Under the coded computation setting, the objective is to design a scheme applicable to any function in a function class (such as multivariate polynomial functions). This motivates developing a suitable notion of locality for a function class, as we do in this section. In order to do so, we define a new notion of locality for an \textit{appropriately defined code} associated with a function class and use this definition to define locality of a function class. We later show that the computational locality of a function class relates directly to its \wt under coded computation (i.e. computational locality is the analog of the number of workers needed for coded computation).

Throughout this section, let $\Fc$ be a class of functions between finite vector spaces $\V = \{v_1,\ldots,v_n\}$ and $\Uu$ over a finite field $\F_q$ and $s$ be a positive integer.

To begin, a codeword is defined for any function.

\begin{definition}[Associated codeword]
\label{def:assCodeword}
For a function $f$ over domain $\V$, the \textbf{associated codeword} is 

$c^f = (f(v_1),\ldots,f(v_n))$.
\end{definition}

Existing literature on locality of codes \cite{locally-decodable-codes,gopalan2012locality,papailiopoulos2014locally,
tamo2014family,prakash2014codes,ramakrishnan2018taking} usually considers a code where each code symbol appears exactly once. The associated codeword reflects this convention. In contrast, the coded computation domain admits replicating a computation over multiple workers, motivating the need for the below definition.

\begin{definition}[Repeated codeword]
For a codeword $c=(c_1,\ldots,c_n) \in \Uu^n$, the \textbf{repeated codeword} is 

$c^\repeatedExponent = ((c_1,1),\ldots,(c_n,1), \ldots, (c_1,s+1),\ldots,(c_n,s+1)).$
\label{def:repCodeword}
\end{definition}

The repeated codeword $c^\repeatedExponent$ repeats the codeword $(s+1)$ times and appends to each repeated code symbol one label from $[s+1]$ for clarity of exposition.

Next, the above definitions are extended to apply to a function class. 
Specifically, Definition~\ref{def:assCode} is used to define a code associated to the function class. 
\begin{definition}[Associated code]
\label{def:assCode}
For a function class $\Fc$, the \textbf{associated code} is $C^{\Fc} = \{c^f \mid f \in \Fc\}$.
\end{definition}

Definition~\ref{def:repCode} generalizes the notion of a repeated codeword to a code.

\begin{definition}[Repeated code]
\label{def:repCode}
For a code $C$, the \textbf{repeated code} is $C^\repeatedExponent = \{c^\repeatedExponent \mid c \in C\}$.
\end{definition}

The computational locality of code symbols is defined and then used to define the computational locality of a code. The definition of computational locality of a code will then be used to define the relevant notion of locality for a function class.

\begin{definition}[\textit{Computational locality of code symbols}]
\label{def:compLocSym}
Let $\K \le n$ and $s$ be non-negative integers and $I=\{i_1,\ldots,i_\K\} \subseteq [n]$. The \textbf{computational locality} $L_{I,s}$ of code symbols $C_I$ for code $C \subseteq \Uu^n$ is the minimum integer for which there exists a size $L_{I,s}$ set $J \subseteq [(s+1)n]$ such that $\forall c,c' \in C^\repeatedExponent, c_J' \in B(c_J,s) \Rightarrow c_I' = c_I$.
\end{definition}
At a high level, Definition~\ref{def:compLocSym} says that the code symbols $C_I$ can be uniquely decoded using any $L_{I,s}-s$ code symbols of a length $L_{I,s}$ punctured code of $C^\repeatedExponent$. Definition~\ref{def:compLocSym} naturally yields a definition for computational locality for a code itself.

\begin{definition}[\textit{Computational locality of a code}]
\label{def:compLoc}
Let $\K \le n$ and $s$ be non-negative integers. The \textbf{computational locality} $L_{\K,s}$ of a code $C \subseteq \Uu^n$ is $max_{I=\{i_1,\ldots,i_\K\} \subseteq [n]}(L_{I,s})$.
\end{definition}

Finally, the computational locality of a function class is defined in terms of Definition~\ref{def:compLoc}.

\begin{definition}[\textit{Computational locality of a function class}]
\label{def:compLocFunc}
Let $\K \le n$ and $s$ be non-negative integers. The function class $\Fc$ has computational locality $L_{\K,s}(\Fc)$ where $L_{\K,s}(\Fc)$ is the computational locality of the associated code $C^{\Fc}$.
\end{definition}

Under Definition~\ref{def:compLocFunc}, for a function class $\Fc$ and any input code symbols, one may consider an arbitrary punctured code of the repeated code. The analog of this property under coded computation is that when designing coded computation schemes, for any specific $\K$ input points one can query the workers at arbitrary evaluation points. This flexibility with queries directly results in a lower \wt under the locality-based model for coded computation than is possible in the \inob setting for multivariate polynomial functions.

\subsection{Connection Between Computational Locality and Coded Computation}
\label{sec:connectionsLocalComp}

Below, it is shown that the computational locality of a function class is equivalent to its \wt (i.e. the minimum necessary number of workers) under coded computation. The proof follows from the above definitions and is included for completeness.

\begin{theorem}[\textit{Coded computation via computational locality}]
\label{thm:codedCompVCompLoc}
Let $\K \le |\V|=n$ and $s$ be non-negative integers. The computational locality of the function class $\Fc$ equals the $(\Fc,\K,s)$-\wt (i.e. $L_{\K,s}(\Fc)=\workers_{\Fc,\K,s}$).
\end{theorem}

\begin{IEEEproof}Consider any input $\langle X_1,\ldots,X_\K\rangle$ and related size $\K$ set $I \subseteq [n]$, where $c_I^f = \langle f(X_1),\ldots,f(X_\K)\rangle$. Every set $J \subseteq [(s+1)n]$ (of some size $j$) corresponds to $\langle \widetilde{X_1},\ldots,\widetilde{X_{j}}\rangle$ such that $\langle f(\widetilde{X_1}),\ldots,f(\widetilde{X_{j}})\rangle =(c^f)^\repeatedExponent_J$. As each worker computes exactly one code symbol, the number of workers needed for coded computation equals the number of code symbols required to uniquely decode $c^f_I$ when missing up to $s$ worker outputs (respectively code symbols). This holds for any $\K$ input points, hence the maximum over all input points.

\end{IEEEproof}

A simple upper bound on the computational locality of a function class follows from the $(s+1)$-wise repeating of code symbols.
\begin{remark}
\label{rem:replic}
The computational locality $L_{\K,s}(\Fc)$ is at most $\K(s+1)$.
\end{remark}

\section{Computational Locality of Multivariate Polynomials}
\label{sec:compLocMultPoly}

In this section, we discuss how to adapt existing locality results into the coded computation setting. 
We demonstrate how to design coded computation schemes for degree bounded multivariate polynomials by using \localRecovery schemes for the associated Reed-Muller code. 
In so doing, we reproduce the best known coded computation scheme for multivariate polynomial functions \cite{yu2019lagrange}.

The following notation is used in this section. Let function class $\Fc=\F_q[x_1,\ldots,x_m]$ be the set of all multivariate polynomials of total degree at most $d$ for a finite field $\F_q = \{\alpha_1,\ldots,\alpha_q\}$. Let $f \in \Fc$ and consider non-negative integers $\K$ and $s$. By definition, the associated code $C^{\Fc}$ is a Reed-Muller code \cite{muller1954application,reed1954class}.

Reed-Muller codes are well-studied and exhibit many useful algebraic properties. The most relevant such property to this work is that for any univariate polynomial $p: \F_q \rightarrow \F_q^m$, the corresponding vector $\left(f(p(\alpha_1)), \ldots,f(p(\alpha_q))\right)$ is a puncturing of $c^f$. Moreover, the composition $f(p(z))$ is a degree $d'=deg(f)deg(p)$ univariate polynomial. One can recover a degree $d'$ univariate polynomial via polynomial interpolation of $d'+1$ distinct points. 
Prior works in the literature on the locality of codes have exploited this property in designing \localRecovery schemes for Reed-Muller codes\cite{lipton1990efficient,beaver1990hiding,gemmell1991self,locally-decodable-codes,gemmell1992highly}. We next discuss such existing \localRecovery schemes for Reed-Muller codes and use them to prove a computational locality result for multivariate polynomial functions. At a high level, we show that every set of $\K$ code symbols is part of a corresponding group of $(\K-1)d+s+1$ code symbols for which every group of $s$ symbols is redundant. This requires the added restriction of a sufficiently large field size (i.e. $(\K-1)d+s+1 \le q$), which we will assume henceforth. 

\pgfplotsset{width=.5\textwidth,height=.25\textwidth,compat=1.10}
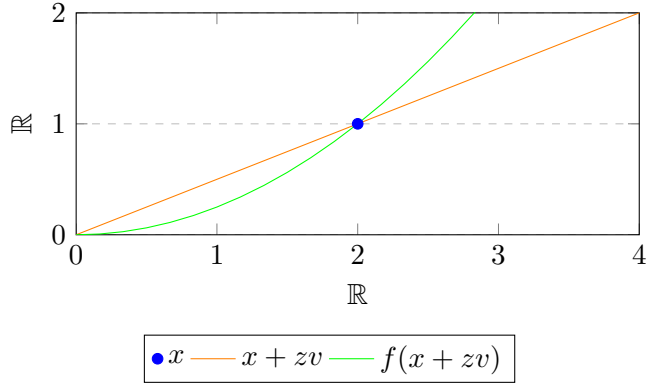
\begin{figure}
\centering

\begin{tikzpicture}[scale=1]
 \tikzstyle{every node}=[]
\centering
\begin{axis}[
    title={},
    xlabel={$\mathbb{R}$},
    ylabel={$\mathbb{R}$},
    xmin=0, xmax=4,
    ymin=0, ymax=2,
    xtick={0,1,2,3,4},
    ytick={0,1,2},
    ymajorgrids=true,
    grid style=dashed,
    legend style={at={(0.45,-0.45)},
anchor=north,legend columns=3},
]
\addplot[
    color=blue,
    mark=*,
    only marks
    ]
    coordinates {
    (2,1)
    };
\addlegendentry{$x$}
  \addplot [
    domain=0:4, 
    color=orange,
    ]
    {.5*x};
\addlegendentry{$x+zv$}
  \addplot [
    domain=0:4, 
    color=green,
    ]
    {(.5*x)*(.5*x)};
\addlegendentry{$f(x+zv)$}

\end{axis}

\end{tikzpicture}

\caption{Example of a \localRecovery scheme for the degree $2$ multivariate polynomial $f(z_1,z_2) = .5z_1z_2$ at $x=(2,1)$. The line $x+zv$ is considered for $z \in \mathbb{R}$ and $v=(4,2).$ The univariate polynomial $f(x+zv)$ is interpolated using $4$ distinct queries tolerating up to $1$ straggling query.}

\label{fig:localEx1}
\end{figure}

Variants of the following \localRecovery scheme for Reed-Muller codes were considered by several previous works \cite{lipton1990efficient,beaver1990hiding,gemmell1991self,locally-decodable-codes}. To recover the value of a polynomial $f$ at a point, $f(x)$, consider the affine line $P = \{x+z v \mid z \in \F_q\}$ for some $v \neq 0 \in \F_q^m$. Let $h(z) = f(x+zv)$ be the degree at most $d$ univariate polynomial restriction of $f$ to $P$ with the property that $h(0) = f(x)$. One can interpolate $h(z)$ by querying it along any $d+s+1$ distinct evaluation points when at most $s$ queries straggle (i.e. are lost). 
A simple example of this scheme is demonstrated in Figure~\ref{fig:localEx1} for the specific polynomial $f(z_1,z_2) = .5z_1z_2$ for $z_1,z_2 \in \mathbb{R}$.

Another \localRecovery scheme follows a similar approach using using a degree $2$ curve instead of an affine line \cite{gemmell1992highly}. Consider the curve $P = \{x + z v_1 + z^2 v_2 \mid z \in \F_q\}$ for $v_1,v_2\neq 0 \in \F_q^m$. Let $h(z) = f(x + zv_1+z^2v_2)$ be the degree at most $2d$ univariate polynomial restriction of $f$ to $P$ such that $h(0) = f(x)$. Interpolation of $h(z)$ follows immediately from querying $h(z)$ at any $2d+s+1$ distinct evaluation points subject to at most $s$ straggling queries.
 
The above \localRecovery schemes were designed to recover the value of a polynomial $f$ at a single input point $x$. Yet the computational locality setting requires evaluating $f$ at multiple input points. A standard generalization of the above schemes using analogous technique with a parameterized degree $d'$ curve is simple to convert to the computational locality regime. Under the generalized \localRecovery scheme, to recover $f(x)$, one can consider the degree $d'$ curve $P = \{x + \sum_{j=1}^{d'} z^j v_j \mid z \in \F_q\}$ for $v_1,\ldots,v_{d'-1},v_{d'} \neq 0 \in \F_q^m$. Let $h(z) = f(x + \sum_{j=1}^{d'} z^j v_j)$ be the degree at most $d'd$ univariate polynomial restriction of $f$ to $P$ such that $h(0) = f(x)$. One can interpolate $h(z)$ by querying it along $d'd+s+1$ distinct evaluation points tolerating up to $s$ straggling queries.

\pgfplotsset{width=.5\textwidth,height=.25\textwidth,compat=1.10}
\begin{figure}
\centering

\begin{tikzpicture}[scale=1]
 \tikzstyle{every node}=[]
\centering
\begin{axis}[
    title={},
    xlabel={$\mathbb{R}$},
    ylabel={$\mathbb{R}$},
    xmin=0, xmax=4,
    ymin=0, ymax=2,
    xtick={0,1,2,3,4},
    ytick={0,1,2},
    ymajorgrids=true,
    grid style=dashed,
    legend style={at={(0.45,-0.45)},
anchor=north,legend columns=3},
]
\addplot[
    color=blue,
    mark=*,
    only marks
    ]
    coordinates {
    (0,0)(2,1)(4,0)
    };
\addlegendentry{$\langle x_1,x_2,x_3\rangle$}
  \addplot [
    domain=0:4, 
    color=orange,
    ]
    {(x-2)*(x-2)*(-1)*.25+1};
\addlegendentry{$p(z)$}
  \addplot [
    domain=0:4, 
    color=green,
    ]
    {((x-2)*(x-2)*(-1)*.25+1)*.5*x};
\addlegendentry{$f(p(z))$}

\end{axis}

\end{tikzpicture}

\caption{Example of a \localRecovery scheme for the degree $2$ multivariate polynomial $f(z_1,z_2) = .5z_1z_2$ at $\langle x_1,x_2,x_3\rangle = \langle (0,0), (2,1), (4,0) \rangle$. The degree $2$ curve $p(z)=(z,1-\frac{(z-2)(z-2)}{4})$ is considered for $z \in \mathbb{R}$. This curve intersects $x_1$ at $z=0$, $x_2$ at $z=2$, and $x_3$ at $z=4$. The univariate polynomial $f(p(z))$ is interpolated using any $5$ out of $6$ distinct queries.}

\label{fig:localEx2}
\end{figure}
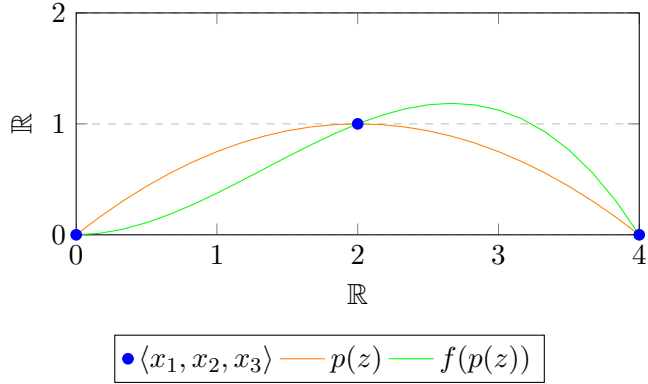

One can adapt the above generalized \localRecovery scheme for Reed-Muller codes, via the locality-based model, into a result on the computational locality of $\Fc$. For any $\K$ input points $\langle X_1,\ldots,X_\K \rangle$, there is a degree $d' \le \K-1$ curve, $p(z)$, along with $\alpha_1,\ldots,\alpha_\K \in \F_q$, so that $p(\alpha_i) = X_i$ for all $1 \le i \le \K$. Thus, the \localRecovery scheme corresponding to $P_{X_1,\ldots,X_\K} = \{p(z) \mid z \in \F_q\}$ computes $h(z) = f(p(z))$ subject to $s$ stragglers; furthermore, $\langle h(\alpha_1),\ldots,h(\alpha_\K) \rangle = \langle f(X_1),\ldots,f(X_\K) \rangle.$ 
A simple illustration of this scheme is shown in Figure~\ref{fig:localEx2} for the polynomial $f(z_1,z_2) = .5z_1z_2$ for $z_1,z_2 \in \mathbb{R}$. 
The aforementioned \localRecovery scheme demonstrates that $\Fc$ has computational locality $L_{\K,s}(\Fc) \le (\K-1)d+s+1$. Combining this bound with that of $(s+1)-$wise replication observed in Remark~\ref{rem:replic} leads to the following result.

\begin{theorem}
\label{thm:compLoc}
The computational locality for total degree at most $d$ multivariate polynomials is $L_{\K,s}(\Fc) \le min(\K(s+1), (\K-1)d+s+1)$. 
\end{theorem}

Although the focus of this section was on multivariate polynomial functions in $\Fc$ from $\F_q^m$ to $\F_q$, the techniques used here easily extend, through component-wise application, to the setting of Cartesian products of functions in $\Fc$.

\subsection{Computational Locality-Based Interpretation of Existing Work}

The LCC scheme \cite{yu2019lagrange} computes the value of $f$, the Cartesian product of any multivariate polynomials, at $\K$ points tolerating up to $s$ stragglers. The scheme either (1) replicates each computation over $(s+1)$ workers or (2) computes the univariate polynomial $f(p(z))$ where $p(z)$ is the Cartesian product of degree at most $\K-1$ polynomials passing through the $\K$ input points. This follows from applying component-wise polynomial interpolation to $(\K-1)deg(f)+s+1$ distinct evaluation points of $f(p(z))$. Thus, the LCC scheme can be viewed as the analogue of the final Reed-Muller \localRecovery scheme described in Section~\ref{sec:compLocMultPoly} (combined with $(s+1)-$wise replication).

\subsection{Computational Locality with Corruptions} 
\label{sec:corrupt}
We next consider the notion of a ``computational locality with corruptions'' of a function class $\Fc$ over $\K$ input points which incorporates up to $\byz$ corrupt in addition to $s$ straggling queries. Computational locality with corruptions is analogous to a coded computation setting containing up to $\byz$ byzantine (adversarial) as well as $s$ straggling servers. 

When $\Fc$ is a class of total degree at most $d$ multivariate polynomial functions, it is simple to adapt the aforementioned \localRecovery schemes to apply to the new setting. For example, one can still compute the values of any $f \in \Fc$ at $\K$ points by obtaining the univariate polynomial restriction of $f$ to $P$, where $P$ is a degree at most $\K-1$ curve passing through all $\K$ input points. To do so, one can use the Berlekamp-Welsh algorithm on $(\K-1)d+2\byz+1$ received (non-straggling) points, rather than interpolating $(\K-1)d+1$ received points with no corruptions \cite{welch1986error}. Moreover, one can adapt the simple $(s+1)$-wise replication technique to such a setting by replicating each input instead $(s+2\byz+1)$ times and using majority-based decoding.

\begin{theorem}
\label{thm:genCCO}
The function class $\Fc$ of total degree at most $d$ multivariate polynomials for $\K$ input points, $\byz$ byzantine corruptions, and $s$ stragglers, has a \wt of at most $ min(\K(s+2\byz+1),(\K-1)d+s+2\byz+1)$.

\end{theorem}

\section{Improved \WT via Locality}

\label{sec:Local}

In Section~\ref{sec:compLocMultPoly}, we have discussed the locality properties of multivariate polynomials at arbitrary input points without using the structure of the specific input points. The results from \cite{yu2019lagrange} show that the overhead in the number of workers is high for coded computation schemes which likewise are oblivious to the structure of the input points. Specifically, when tolerating up to $s$ stragglers for a polynomial of total degree at least $(s+1)$, it is impossible to outperform the factor $s$ overhead on the number of workers required by $(s+1)-$wise replication.

The focus of this section is how the locality-based approach to model coded computation aids in going beyond the \wt of the \inob setting. This improvement follows from enabling schemes to adaptively exploit the structure of specific input points.  
In Section~\ref{sec:homogLoc}, we show how to leverage {linear dependencies} in the input points to reduce the required number of workers for coded computation of a homogeneous multivariate polynomial function. In Section~\ref{sec:nonhomogPoly}, we use homogenization\textemdash a technique by which one converts a multivariate polynomial into a homogeneous multivariate polynomial\textemdash to likewise exploit linear dependencies in the input points for non-homogeneous polynomial functions. This will establish that under the proposed locality-based model, the \wt is less than that of $(s+1)$-wise replication which is necessary in the \inob setting. Moreover, in Section~\ref{sec:sparseLD}, we demonstrate concrete settings in which this improvement is significant. 
In Section~\ref{sec:localClass}, we discuss the rich class of locality properties that coded computation schemes can exploit enabled by the proposed locality-based model. 
Finally, in Section~\ref{sec:adversarialLocal}, we extend all of our results to a setting in which there are at most $\byz$ byzantine (adversarial) workers in addition to up to $s$ straggling workers.

\subsection{Coded Computation of Homogeneous Multivariate Polynomials}
\label{sec:homogLoc}

In this section, we focus on the class of homogeneous multivariate polynomial functions. Such polynomials have the fundamental property that multiplying the input by a field element is equivalent to multiplying the output by an associated power of the field element. We exploit this property in designing a novel coded computation scheme through leveraging locality (in the form of linear dependencies) of the input points.

A homogeneous polynomial $f \in \F[x_1,\ldots,x_m]$ is defined as $f(X) = \sum_{i \in I} p_i(X)$ for some index set $I$ where for each $i \in I$, $p_i \in \F[x_1,\ldots,x_m]$ is a monomial with $deg(p_i) = deg(f)$. Consequently, for any homogeneous polynomial $f$ and any $\alpha \in F$, $f(\alpha X) = \alpha^{deg(f)} f(X).$

\pgfplotsset{width=.5\textwidth,height=.25\textwidth,compat=1.10}
\begin{figure}
\centering
\begin{tikzpicture}[scale=1]
 \tikzstyle{every node}=[]
\centering
\begin{axis}[
    title={},
    xlabel={$\mathbb{R}$},
    ylabel={$\mathbb{R}$},
    xmin=0, xmax=4,
    ymin=0, ymax=2,
    xtick={0,1,2,3,4},
    ytick={0,1,2},
    ymajorgrids=true,
    grid style=dashed,
    legend style={at={(0.45,-0.45)},
anchor=north,legend columns=2},
]
\addplot[
    color=red,
    mark=*,
    only marks
    ]
    coordinates {
    (0,1)(2,0)(2,1)
    };

\addlegendentry{Input points}
 
\addplot[
    color=blue,
    mark=*
    ]
    coordinates {
    (0,2)(4,0)
    };
\addlegendentry{Multiples of input points}
  \addplot [
    domain=0:4, 
    color=blue,
    ]
    {2-x/2};
\addlegendentry{$\lin(z)=(z,2-z/2)$}

\end{axis}

\end{tikzpicture}

\caption{Example for locality-based coded computation of a homogeneous polynomial, $f$. Computing $f(\lin(z))$ determines the value of $f$ at all three input points due to homogeneity.
}

\label{fig:homogEx}
\end{figure}
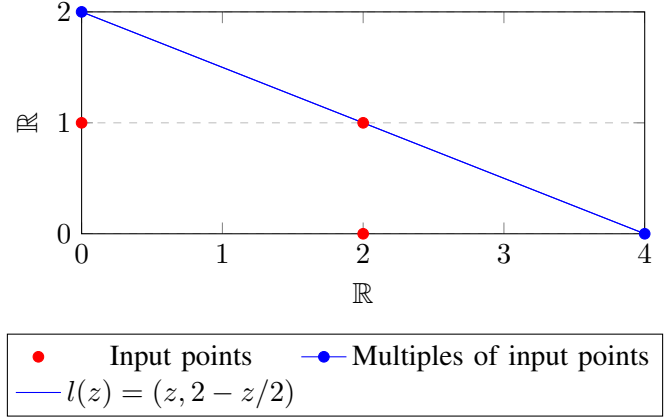Thus, in order to compute the value of $f$ on any input points, it suffices to compute the value of $f$ at \textit{nonzero multiples of the input points}. One can exploit this property to design coded computation schemes using fewer workers than is possible in the \inob setting for linearly dependent input points. We summarize how a coded computation scheme can do so, discuss a simple example of such a scheme, and finally present an explicit construction in full detail. 

Consider any minimal linearly dependent set of $\K$ vectors (i.e. a set of linearly dependent vectors for which no strict subset is linearly dependent). There is a $p(z)$ which is the Cartesian product of degree at most $\K-2$ univariate polynomials passing through a nonzero multiple of each of the $\K$ vectors. Moreover, $f(p(z))$ is the Cartesian product of degree at most $deg(f)(\K-2)$ univariate polynomials and can be interpolated (component-wise) using any $deg(f)(\K-2)+1$ distinct evaluations. Hence, a coded computation scheme can compute $f(p(z))$ and thereby obtain the values of $f$ at nonzero multiples of the $\K$ input points. By homogeneity, this is sufficient to decode $f$ at the $\K$ input points. In contrast, a polynomial $p'(z)$ passing through the $\K$ specific points can be of degree $\K-1$; a similar scheme computing $f(p'(z))$ would require $deg(f)(\K-1)+1$ distinct evaluations of $f(p'(z))$ to be received for interpolation.

In Figure~\ref{fig:homogEx}, we discuss a simple example of a coded computation scheme for a homogeneous polynomial $f$ subject to $s$ straggling workers where $2 \le (s+1) = deg(f)$. The example contains three distinct linearly dependent input points in the x-y plane, $\langle (0,1),(2,0),(2,1)\rangle$. The line $\lin(z)$ intersects a nonzero multiple of each input, and by homogeneity, $f(2 X) = 2^{deg(f)} f(X)$ for $\alpha \in \mathbb{R}$. Hence, computing $f(\lin(z))$ is sufficient to determine $\langle f((0,1)), f((2,0)), f((2,1))\rangle$; this requires $2(s+1)$ workers, while the \wt in the \inob setting is $3(s+1).$

We now present the \textbf{\hs} for any $\K$ minimal linearly dependent input points. The scheme is systematic and uses linear encoding and decoding.

\textbf{Input}: $\langle X_1,\ldots,X_\K \rangle$ where $X_j \in \F^m$ for $1 \le j \le \K$ and $\sum_{i=1}^{\K-1} \alpha_i X_i = X_\K$ for $\alpha_1,\ldots,\alpha_{\K-1} \in \F \setminus \{0\}$ for a field $\F$. A positive integer $s$ of straggling workers and $deg(f),$ the degree of the polynomial to be computed.

\textbf{Evaluation points}: Let the first $\K$ evaluation points be $\widetilde{X_i} = X_i$ for $1 \le i \le \K$. 
Let $\beta_1,\ldots,\beta_{(\K-2)deg(f)+s+1} \in F$ be distinct elements and $p^*(z) = \sum_{i=1}^{\K-1} \alpha_i X_i.\prod_{j \in [\K-1] \setminus \{i\}} (z-\beta_j)  (\beta_\K-\beta_j)^{-1}.$ 
Let the remaining evaluation points be $\widetilde{X_i} = p^*(\beta_i)$ for $\K+1 \le i \le (\K-2)deg(f)+s+1$. The scheme queries $f(\widetilde{X_i})$ for $1 \le i \le (\K-2)deg(f)+s+1$.

\textbf{Decoding}: For each received worker output $i \in [\K-1]$ compute 
$f(p^*(\beta_i)) = f(\widetilde{X_i}) \cdot \left( \alpha_i\prod_{j \in [\K-1] \setminus \{i\}}\frac{\beta_i-\beta_j}{\beta_\K-\beta_j}  \right)^{deg(f)}.$ As $f(\widetilde{X_i})=f(p^*(\beta_i))$ for $\K \le i \le (\K-2)deg(f)+s+1$, the scheme now has access to $(\K-2)deg(f)+1$ distinct evaluations of $f(p^*(z))$. 
It uses them to interpolate $f(p^*(z))$. 
Decoding follows from setting $f(X_\K)=f(p^*(\beta_\K))$ and for $i \in [\K-1]$, setting $f(X_i)=f(p^*(\beta_i)) \cdot\left( \alpha_i\prod_{j \in [\K-1] \setminus \{i\}}\frac{\beta_i-\beta_j}{\beta_\K-\beta_j} \right)^{-deg(f)}.$

Correctness of the \hs is shown in Theorem~\ref{thm:homogeneous}. 

\begin{theorem}
\label{thm:homogeneous}
Let $\langle X_1,\ldots,X_\K \rangle$ be $\K$ input points over $\F^m$ for field $\F$ such that $\sum_{i=1}^{\K-1} \alpha_i X_i =X_\K$ for $\alpha_1,\ldots,\alpha_{\K-1} \in \F \setminus \{0\}$. Let $s$ be a non-negative integer and $|\F| \ge (\K-2)deg(f)+s+1$. Then the \hs computes $\langle f(X_1),\ldots, f(X_\K)\rangle$ tolerating up to $s$ straggling workers using $(\K-2)deg(f)+s+1$ workers.
\end{theorem}

\begin{IEEEproof}Let $\lambda_i = \alpha_i \prod_{j \in [\K-1] \setminus \{i\}}(\beta_i-\beta_j)(\beta_\K-\beta_j)^{-1}$
and note that $\lambda_i \in \F \setminus \{0\}$ for $1 \le i < \K.$ Thus, $p^*(\beta_\K) = \sum_{i=1}^{\K-1}\alpha_i X_i = X_\K$ and for $1 \le i < \K$ $p^*(\beta_i) = X_i\lambda_i.$ Due to  the homogeneity of $f$, this means $f(p^*(\beta_i)) = \lambda_i^{deg(f)} f(X_i)$ for $1\le i < \K$ and  $f(p^*(\beta_\K)) = f(X_\K)$.

As $deg(p^*(z)) \le \K-2$, one can interpolate $f(p^*(z))$ with any $(\K-2)deg(f)+s+1$ distinct evaluations of $f(p^*(z))$, while tolerating up to $s$ stragglers. One can decode $\langle f(X_1),\ldots,f(X_k)\rangle$ using $f(p^*(z)).$

\end{IEEEproof}

One can apply the \hs to any $\K>m$ input points over $\F^m$ by using Gaussian Elimination to partition the input points into $\parti_1,\ldots,\parti_\partisize,$ and $E$. Each $\parti_i$ for $1 \le i \le \partisize$ is comprised of a set of minimal linearly dependent input points where $\partisize \ge \lfloor \frac{\K}{m+1} \rfloor$ and $E$ contains the at most $m$ remaining input points. One can apply the \hs to each of $\parti_1,\ldots,\parti_\partisize$ and $(s+1)$-wise replication to each input point in $E$, provided the field size is at least $(m-1)deg(f)+s+1$. When $deg(f) =s+1$, this scheme combining \hs with $(s+1)-$wise replication requires at most $(\K-\partisize)(s+1)$ workers. In contrast, the \wt under coded computation of the \inob setting is $\K(s+1)$. This leads to the following result.

\begin{theorem}[Improved \wt for homogeneous polynomials]
Let $\Fc$ be the function class a total degree $(s+1)$ homogeneous polynomial over domain $\F^m$. For $\K$ input points where $(m+1)|\K$ and at most $s$ stragglers, the \wt under coded computation in the locality-based model is at most $\K \frac{m}{m+1} (s+1)$.
\label{thm:homogFull}
\end{theorem}

\subsection{Coded Computation of Non-Homogeneous Multivariate Polynomials}

\label{sec:nonhomogPoly}
We have shown for homogeneous multivariate polynomial functions that the \wt under the locality-based model for coded computation is lower than that of the \inob setting. In this section, we extend this result to non-homogeneous multivariate polynomial functions by combining the \hs and the technique of homogenizing polynomials.

We begin by reviewing homogenizing polynomials. Let $f \in \F[x_1,\ldots,x_m]$ be a multivariate polynomial function for a field $\F$. The homogenizing polynomial of $f(x_1,\ldots,x_m)$ is defined as $f'(r,(x_1,\ldots,x_m)) = r^{deg(f)} f(x_1r^{-1},\ldots,x_mr^{-1})$. Furthermore, $f'$ is homogeneous and $f'(1,(x_1,\ldots,x_m)) = f(x_1,\ldots,x_m)$. The point $(1,(x_1,\ldots,x_m))$ is denoted as the ``homogenizing'' point of $(x_1,\ldots,x_m)$. 

We now present the \textbf{\ns}. The scheme works by converting the input points into homogenizing points and then applying the \hs with the homogenizing polynomial $f'$ using as input points the homogenizing points.

\textbf{Input}: $\langle X_1,\ldots,X_\K \rangle$ where each $X_j \in \F^m$ for $1 \le i \le \K$ and  $\sum_{i=1}^{\K-1} \alpha_i (1,X_i) = (1,X_\K)$ for $\alpha_1,\ldots,\alpha_{\K-1} \in \F \setminus \{0\}$ for a field $\F$. A non-negative integer $s$ of straggling workers and $deg(f),$ the degree of the non-homogeneous polynomial.

\textbf{Evaluation points}: Identify the evaluation points $\widetilde{X_i}'=(r_i,X_i^*) \in \F^{m+1}$ for $i \in [(\K-2)deg(f)+s+1]$ used by the \hs on the inputs of homogenizing points $\langle X_1',\ldots,X_\K' \rangle = \langle (1,X_{1}),\ldots,(1,X_{\K}) \rangle,$  non-negative integer $s$, and $deg(f)$. Without loss of generality, do so with each $r_i \neq 0$. The scheme queries $f(\widetilde{X_i}) = f(r_i^{-1}X_i^*)$ for $i \in [(\K-2)deg(f)+s+1]$.

\textbf{Decoding}: Multiply each received $f(\widetilde{X_i}) = f(r_i^{-1}X_i^*),$ by $r_i^{deg(f)}$ to determine the value of $f'$ at the corresponding evaluation points, namely $f'(\widetilde{X_i'}) = f'(r_i,\widetilde{X_i}) = r_i^{deg(f)}f(r_i^{-1} r_i^{-1}X_i^*)$. Complete the \hs decoding to determine $\langle f'((1,X_1)),\ldots,f'((1,X_\K)) \rangle = \langle f(X_1),\ldots,f(X_\K) \rangle.$

Correctness of the \ns is verified in Theorem~\ref{thm:nonhomogeneousFullScheme}.

\begin{theorem}
\label{thm:nonhomogeneousFullScheme}
Let $\langle X_1,\ldots,X_\K \rangle$ be $\K$ input points over $\F^m$ for field $\F$ for which $\sum_{i=1}^{\K-1} \alpha_i (1,X_i) =(1,X_\K)$ for $\alpha_1,\ldots,\alpha_{\K-1} \in \F \setminus \{0\}$. Let $s$ be a non-negative integer and $|\F| \ge (\K-2)(deg(f)+1)+s+1$. Then the \ns computes $\langle f(X_1),\ldots, f(X_\K)\rangle$ tolerating up to $s$ straggling workers using $(\K-2)deg(f)+s+1$ workers.
\end{theorem}

\begin{IEEEproof}Due to Theorem~\ref{thm:homogeneous} and properties of homogenizing polynomials, we need only prove that there are enough distinct queries of each low degree curve so that each query of $f'(\widetilde{X_i'}) = f'(r_i,X_i^*)$ has $r_i \neq 0$ for $i \in [(\K-2)deg(f)+s+1]$.

Recall from the proof of Theorem~\ref{thm:homogeneous} that the evaluation points $\widetilde{X_i}$ for $i \in [(\K-2)deg(f)+s+1]$ lie on the curve $p^*(z)$ (i.e. $\widetilde{X_i} = p^*(\beta_i)$ for $\beta_i \in \F$). Moreover, for $i \in [\K]$, the evaluation point $\widetilde{X_i} = p^*(\beta_i)= (1,X_i)$, the homogenizing point for the input point $X_i$. Therefore, the first coordinate of the output of $p^*(z)$ can only equal $0$ for at most $deg(p^*(z))$ inputs. 
Since $deg(p^*(z)) \le \K-2$, the \ns requires at most $(\K-2)deg(f)+s+1$ evaluation points. Moreover, it discards at most $\K-2$ points where $p^*(z)$ is $0$. This is feasible, as $|\F| \ge (\K-2)(deg(f)+1)+s+1$. 
\end{IEEEproof}

Similar to extending the \hs to $\K>m$ arbitrary input points in the homogeneous setting, one can extend the \ns to apply to $\K>m+1$ arbitrary input points over $\F^m$ in the non-homogeneous setting. To do so, one partitions the input points $\langle X_1,\ldots,X_\K \rangle$ into $\parti_1,\ldots,\parti_\partisize,$ and $E$ using Gaussian Elimination. Each $\parti_i$ for $1 \le i \le \partisize$ consists of $\ls_i \le m+2$ input points $\langle X^{(i)}_{1},\ldots,X^{(i)}_{\ls_i}\rangle$ such that the corresponding set of homogenizing points $\{(1,X^{(i)}_{1}),\ldots,(1,X^{(i)}_{\ls_i})\}$ is minimally linearly dependent. Moreover, $E$ contains the at most $m+1$ extra input points, and $\partisize \ge \lfloor \frac{K}{m+2}\rfloor$. One can apply the \ns to each of $\parti_1,\ldots,\parti_\partisize$ and $(s+1)$-wise replication to each input point in $E$, provided the field size is at least $m(deg(f)+1)+s+1$. Overall, this requires at most $(\K-\partisize)(s+1)$ workers when $deg(f)=(s+1)$. In contrast, the \wt in the \inob setting for coded computation is $\K(s+1)$. This leads to the following result.

\begin{theorem}[Improved \wt for non-homogeneous polynomials]
\label{thm:nonHomResource}
Let $\Fc$ be the function class a non-homogeneous multivariate polynomial of total degree $(s+1)$ over domain $\F^m$. For $\K$ input points where $(m+2)|\K$ and up to $s$ stragglers, the \wt under the locality-based model for coded computation is at most $\K \frac{m+1}{m+2}(s+1)$.
\end{theorem}

While the improvement in \wt over that of \inob setting may not be substantial, Theorem~\ref{thm:nonHomResource} shows that the lower bound on \wt from \cite{yu2019lagrange} is not a fundamental barrier for coded computation of non-linear functions. This opens up the potential of more resource efficient coded computation schemes of non-linear functions.

\subsection{Sparse Linear Dependencies in Input Points}

\label{sec:sparseLD}

We have seen in Sections~\ref{sec:homogLoc} and~\ref{sec:nonhomogPoly} coded computation schemes which exhibit improved \wt over the best possible in the \inob setting. The extent of the reduction in the number of workers needed depends on the sizes of minimal linearly dependent sets of input points. When the size of a set of linearly dependent vectors is small, we refer to it as ``sparse'' (i.e. involving few vectors). Under the locality-based model of coded computation, improvement in \wt is considerable when there are sparse linear dependencies in the input points. The fewer the number of input points per linear dependency, the more pronounced the improvement. In this section, we discuss how such sparse linear dependencies in the input points naturally arise in certain applications. 
We then consider the regime where the input points are over a finite vector space $\V = \F_q^m$ over a finite field $\F_q$. 
We show that if there are sufficiently many input points, then sparse linear dependencies in the input points must exist. 
This results in a significantly lower \wt under the locality-based approach for coded computation than the best possible in the \inob setting.

In certain applications, the set of input points is augmented with additional points. These extra points are each comprised of linear combinations of a small number of the original input points. 
For example, in one recently proposed approach to training neural networks for image classification \cite{zhang2017mixup}, the authors propose ``mixup'', wherein the training data is augmented with linear combinations of a small number of points from the original training data. 
For any two training data points, $X_1$ and $X_2$, a new training example $X_1\alpha +(1-\alpha)X_2$ for $0<\alpha <1$ might be added. Under the locality-based model for coded computation, one can exploit such sparse linear dependencies (involving the augmented data points) to reduce the required number of workers.

One can always exploit a set of minimal linearly dependent input points to reduce the \wt for homogeneous multivariate polynomial computations by using the \hs. In contrast, for non-homogeneous multivariate polynomial computations, even for linearly dependent input points $\langle X_1,\ldots,X_\K \rangle$, the corresponding homogenizing points $\langle (1,X_1),\ldots,(1,X_\K) \rangle$ may be linearly independent. However, when a partition of the input points includes the disjoint linearly dependent sets $\{ X_1,\ldots,X_{\K_1} \}$ and $\{ X_1',\ldots,X_{\K_2}' \}$, the set of homogenizing points $\{ (1,X_1),\ldots,(1,X_{\K_1}), (1,X_1'),\ldots,(1,X_{\K_2}') \}$ is linearly dependent. This follows from the fact that either (1) at least one of $\{ (1,X_1),\ldots,(1,X_{\K_1})\}$ and $\{(1,X_1'),\ldots,(1,X_{{\K_2}}') \}$ is linearly dependent or (2) $(1,0,\ldots,0)$ is in their intersection.
Thus, it is always possible to exploit any two minimal linearly dependencies of disjoint input points to reduce the \wt by using the \ns.

In Lemma~\ref{lem:filLinDep}, we show that whenever there are sufficiently many $(\K)$ points, there exists a size $2\ls$ linearly dependent subset of these points. A finite vector space over a finite field $\F_q$ contains a fixed number of vectors, while the number of linear combinations with nonzero coefficients of $\ls$ out of $\K$ input points is $(q-1)^\ls \binom{\K}{\ls}.$ When $(q-1)^\ls \binom{\K}{\ls}$ exceeds the size of the vector space, there exists $2$ distinct linear combinations of $\ls$ input points which are equal. The set of involved vectors in the $2$ linear combinations is therefore linearly dependent.

\begin{lemma}
\label{lem:filLinDep}
For $\ls \ge 2$, a set of $\K> 2\ls q^{\frac{m}{\ls}-1}$ nonzero vectors of $\F_q^m$ must contain a size at most $2\ls$ linearly dependent subset. 
\end{lemma}

\begin{IEEEproof}Let there be $\K$ points $X_1,\ldots,X_\K \in \F_q^m \setminus \{0\}^m$ and consider the $\binom{\K}{\ls}$ sets of $\ls$ points. If any such set is not full rank, the proof is concluded. Otherwise, for each size $\ls$ set $I \subseteq [\K]$ there are $(q-1)^\ls$ distinct vectors of the form $\sum_{i \in I} \alpha_{i,I} X_i$ where $\alpha_{i,I} \in \F \setminus \{0\}$. Overall, there are $\binom{\K}{\ls}(q-1)^\ls$ such vectors. When $\binom{\K}{\ls} (q-1)^\ls>q^m$, there exists two of these vectors which are equal. As $\binom{\K}{\ls} (q-1)^\ls \ge \frac{\K}{\ls}^\ls (q-1)^\ls \ge \frac{\K}{2\ls}^\ls q^\ls > q^m,$ this condition is met when $\K>2\ls q^{\frac{m}{\ls}-1}$. 

 Let $\sum_{i \in I} \alpha_{i,I} X_i = \sum_{j \in J} \alpha_{j,J} X_j$ for $\alpha_{i,I},\alpha_{j,J} \in \F \setminus \{0\}$ and distinct $I,J \subseteq [\K]$. Then $\sum_{i \in I \setminus J}  \alpha_{i,I} X_i - \sum_{j \in J \setminus I} \alpha_{j,J} X_j + \sum_{l \in I \cap J} (\alpha_{l,I}-\alpha_{l,J}) X_l = 0$. Both $I \setminus J$ and $J \setminus I$ are nonempty with all nonzero corresponding coefficients $\alpha_{i,I}$ (for $i \in I \setminus J$) and $\alpha_{j,J}$ (for $j \in J \setminus I$). Hence, there is a linear dependence of at most $2\ls$ terms.

\end{IEEEproof}

When there sufficiently many input points, there must be sparse linear dependencies in the input points by Lemma~\ref{lem:filLinDep}. Consequently, one can partition all but negligibly many of the input points into size at most $2\ls$ linearly dependent sets. Thus, for a homogeneous or non-homogeneous multivariate polynomial $f$ of total degree $(s+1)$, $s$ stragglers, and $\K$ input points, one can apply the \hs or the \ns to the size at most $2\ls$ linearly dependent sets. Combining this with $(s+1)-$wise replication for the remaining input points requires $\approx \K \frac{2\ls-1}{2\ls}(s+1)$ workers in total. In contrast, the \wt in the \inob setting is $\K(s+1)$.

\begin{theorem}
Suppose $\Fc$ be the function class of total degree $(s+1)$ polynomials over domain $\F_q^m$, at most $s$ workers straggle, and $\K> 2\ls q^{\frac{m+1}{\ls}-1}$. The \wt under locality-based coded computation is at most $\K \left( \frac{2\ls-1}{2\ls}(s+1) + (s+1)\frac{2\ls q^{\frac{m+1}{\ls}-1}}{\K} \right)$.
\end{theorem}

\subsection{General Class of Locality-Based Coded Computation Schemes}
\label{sec:localClass}

Under the locality-based model, coded computation schemes can leverage the advantageous structural properties of the specific input points. 
We showed in Sections~\ref{sec:homogLoc} and~\ref{sec:nonhomogPoly} one way to leverage locality properties of input points to design coded computation schemes using fewer workers than is possible in the \inob setting. 
We demonstrate the richness of the class of methods for using locality properties of the input points by presenting two more techniques to exploit locality properties of the input points in Section~\ref{sec:cic}. In Section~\ref{sec:coc}, we discuss how one can use locality properties of the output points (i.e. the image of the input points under the computed function) to reduce the number of workers for coded computation.

\subsubsection{More on Locality in Input Points}

\label{sec:cic}
We now discuss two additional locality properties of input points: First, we consider the simple scenario where input points lie on a particularly low degree curve and observe that the points have a useful locality structure. Specifically, suppose there are $\K$ input points $\langle X_1,\ldots,X_\K\rangle$ over $\F^m$ for a field $\F$ and a degree at most $(\K-2)$ curve $p^*(z)$ such that $p^*(\beta_i) = X_i$ where $\beta_i \in \F$ for $i \in [\K]$. For any $f$ which is the Cartesian product of multivariate polynomial functions, the univariate polynomial $f(p^*(z))$ can be interpolated (component-wise) using any $deg(f)deg(p^*(z))+1 \le deg(f)(\K-2)+1$ distinct evaluations. Thus, the value of $f$ at the $\K$ inputs can be computed while tolerating up to $s$ stragglers using $deg(f)(\K-2)+s+1$ workers each querying a distinct evaluation of $f(p^*(z))$.

Second, we demonstrate a locality property of input points for multivariate polynomials in the following scenario, which we denote ``intersecting curves;'' there are two low degree curves which pass through distinct input points and also intersect each other. We design a scheme to exploit this structure to achieve lower \wt than is possible in the \inob setting. 
At a high level, the scheme interpolates the two low degree curves. Computing one curve is sufficient to compute the intersection point, which can then be used in interpolating the another other curve. The intersection point is used to make do with $2$ fewer queries.

\pgfplotsset{width=.5\textwidth,height=.25\textwidth,compat=1.10}
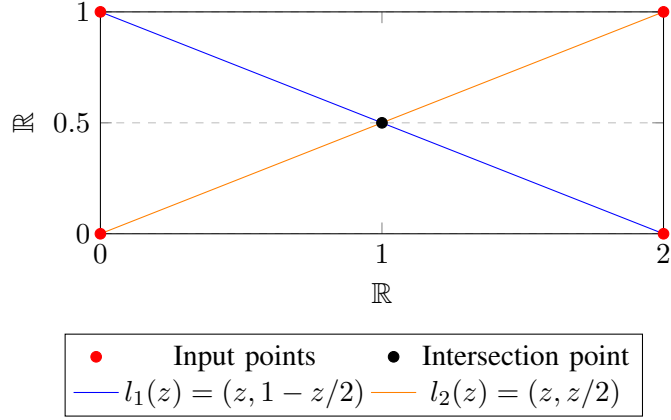
\begin{figure}
\centering
\begin{tikzpicture}[scale=1]
 \tikzstyle{every node}=[]
 \centering
\begin{axis}[
    title={},
    xlabel={$\mathbb{R}$},
    ylabel={$\mathbb{R}$},
    xmin=0, xmax=2,
    ymin=0, ymax=1,
    xtick={0,1,2},
    ytick={0,.5,1},
   legend style={at={(0.45,-0.45)},
anchor=north,legend columns=2},
    ymajorgrids=true,
    grid style=dashed
]
\addplot[
    color=red,
    mark=*,
    only marks
    ]
    coordinates {
    (0,0)(0,1)(2,0)(2,1)
    };
\addlegendentry{Input points}
  \addplot[
    color=black,
    mark=*,
    only marks
    ]
    coordinates {
    (1,.5)
    };
\addlegendentry{Intersection point}

 \addplot [
    domain=0:2, 
    color=blue,
    ]
    {1-x/2};
\addlegendentry{$\lin_1(z)=(z,1-z/2)$}

 \addplot [
    domain=0:2, 
    color=orange,
    ]
    {x/2};
\addlegendentry{$\lin_2(z)=(z,z/2)$}

\end{axis}
\end{tikzpicture}

\caption{Example for locality-based coded computation of a polynomial $f$ at $4$ input points using two intersecting lines. The lines $\lin_1(z)$ and $\lin_2(z)$ pass through the $4$ input points and intersect at $z=1$. Using the identity $f(\lin_1(1)) = f(\lin_2(1))$ reduces the number of workers by $2$.}

\label{fig:collide}
\end{figure}We begin by using a simple example of intersecting curves in Figure~\ref{fig:collide} before describing the scheme in detail. In the example, the lines $\lin_1(z) =(z,1-z/2)$ and $\lin_2(z) = (z,z/2)$ pass through the input points $\{(0,1),(2,0)\}$ and $\{(0,0),(2,1)\}$ while intersecting at $\lin_1(1) = \lin_2(1) = (1,.5)$. After computing $f(\lin_1(z))$, the point $f(\lin_1(1)) = f(\lin_2(1))$ can be recovered and used in computing $f(\lin_2(z))$ (and vice versa). Consequently, querying the workers with evaluation points of $f$ along the two lines and exploiting the intersection point reduces the number of workers needed.

We now present the \textbf{\cics} for any degree $d$ polynomial subject to $s$ stragglers for input points which lie on two low-degree intersecting curves.

\textbf{Input}: Distinct $\langle X_1,\ldots,X_k,X_1',\ldots,X_{k'}' \rangle$ over $\F^m$ for field $\F$ where the lowest degree curves passing through $\langle X_1,\ldots,X_k\rangle$ and $\langle X_1',\ldots,X_{k'} \rangle$ intersect. A non-negative integer $s$ of stragglers and $deg(f)$, the degree of the polynomial to be computed. 

\textbf{Evaluation point}: Let $p(z)$ and $p'(z)$ be the degree less than $\K$ and $\K'$ univariate polynomials such that: (1) $p(\alpha_i) = X_i$ for $\alpha_i \in \F$ and $i \in [\K] $. (2) $p'(\beta_i) = X_i'$ for $\beta_i \in \F$ and $i \in [\K']$. (3) $p(z_*) = p'(z_{*'}')$ for some $z_*,z_{*'}' \in \F$. For $i \in [deg(f) \cdot deg(p(z)) +s]$, let $\widetilde{X_i} = f(p(\alpha_i))$ for distinct $\alpha_i \in \F \setminus \{z_*\}.$ For $deg(f) \cdot deg(p(z)) +s \le i  \le deg(f) \cdot (deg(p(z))+deg(p'(z))) +2s$, let $\widetilde{X_i} = f(p'(\beta_i))$ for distinct $\beta_i \in \F \setminus \{z_{*'}'\}.$ 
Query the $i$th worker at $\widetilde{X_i}$ to compute $f(\widetilde{X_i})$ for $i \in [deg(f) \cdot (deg(p(z))+deg(p'(z))) +2s].$

\textbf{Decoding}: use the received queries to interpolate at least one of $f(p(z))$ or $f(p'(z))$ and thereby recover $f(p(z_*)) = f(p'(z_{*'}'))$. Use the decoded intersection point and received points to interpolate the second curve. Decoding follows from setting $\langle f(X_1),\ldots,f(X_k),f(X_1'),\ldots,f(X_{k'}') \rangle=\langle f(p(\alpha_1)),\ldots,f(p(\alpha_k)), f(p(\beta_1)),\ldots,f(p(\beta_{k'})) \rangle.$

\begin{remark}
Correctness for the \cics immediately follows from univariate polynomial interpolation when $|\F| \ge d(max(k,k')-1)+s+1$.
\end{remark}

In Lemma~\ref{lem:collideLines}, we show that if there are enough input points in a finite domain ensure, then there must exist (1) two intersecting lines passing through $4$ input points or (2) three input points which lie on a line. One can exploit either property to reduce the number of workers in the computation. The result follows from the fact that with enough input points, the number of points on the lines between input points exceeds the total number of points in a finite domain. Hence, there must be lines between input points which intersect, necessitating (1) or (2).

\begin{lemma}
\label{lem:collideLines}
Suppose there are $\K$ distinct input points $\langle X_1,\ldots,X_\K \rangle$ over a finite vector space $\F_q^m$ for $\K>1$ and $q>3$. If $\K > 8q^{\frac{m-1}{2}}$, there exists three input points which lie on a line or there exists four input points which lie on two lines which intersect.
\end{lemma}
\begin{IEEEproof}Consider $\lin_{i,j}(z) = X_i + (X_j-X_i)z$ for $1 \le i <j \le \K$. There are $\frac{\K(\K-1)}{2}$ such lines, each with $q-2$ points $\lin_{i,j}(z)$ for $z \in \F_q \setminus \{0,1\}$. There are $q^m$ points in the domain. When $\frac{\K(\K-1)(q-2)}{2} > q^m$, there exists $\lin_{i,j}(z)$ and $\lin_{i',j'}(z)$ which intersect outside of $\{X_i,X_j,X_{i'},X_{j'}\}$.  If $i \in \{i',j'\}$ or $j \in \{i',j'\}$, the two lines intersect in two distinct points and are the same line, hence three input points lie one a single line. If not, $i,j,i',$ and $j'$ are distinct, concluding the proof.
\end{IEEEproof}

One can combine the \cics and polynomial interpolation along lines passing through $3$ input points (as described in the first paragraph of this section) into the following coded computation scheme for multivariate polynomials. Let $f$ be a multivariate polynomial of total degree $(s+1)$, $s$ be the maximum number of stragglers, and $\langle X_1,\ldots,X_\K \rangle$ be the input points. The scheme partitions the input points $\langle X_1,\ldots,X_\K\rangle$ into $\parti_1,\ldots,\parti_\partisize,\parti_1',\ldots,\parti_{\partisize'}',$ and $E$. Each of $\parti_i$ for $1 \le i \le \partisize$ consists of $4$ input points which lie on two intersecting lines, each of $\parti_i'$ for $1 \le i \le \partisize'$ consists of $3$ input points which lie on a single line $\lin_i(z)$, and $E$ contains the extra input points. For each $\parti_i$ for $1 \le i \le \partisize$, the scheme can apply the \cics by querying each line with $2s+1$ workers outside the intersection point (and using the recovered intersection point in decoding). For each $\parti_i'$ for $1 \le i \le \partisize'$, the scheme can use $deg(f)+s+1$ workers to interpolate $f(\lin_i(z)),$ thereby computing the value of $f$ at the at least $3$ input points. Finally, the scheme can apply $(s+1)-$wise replication to the input points in $E$. When the number of input points is $\K \gg q^{\frac{m-1}{2}}$, by Lemma~\ref{lem:collideLines}, $E$ contains a negligible fraction of the input points. Thus, the scheme requires fewer workers than the \wt in the \inob setting of $\K(s+1)$.

\begin{remark}
Let $f$ be a polynomial with $deg(f) = s+1$, there be at most $s$ stragglers, and the number of input points be $\K > 8q^{\frac{m-1}{2}}$. Then the above scheme requires at most $\K(s+\frac{1}{2}+ (s+1)\frac{8q^{\frac{m-1}{2}}}{\K})$ workers.
\end{remark}

\subsubsection{Locality in Output Points}
\label{sec:coc}

Under the locality-based model, one can exploit locality properties of the output points (i.e. the image of the input points under the computed function) in addition to locality properties of the input points to reduce the number of workers needed. We now demonstrate that restricting the class of multivariate polynomial functions to require additional structure can impose locality properties of the output points. Such a restriction is relevant to applications that inherently do so; for example, considering matrix products rather than arbitrary polynomials in distributed matrix multiplication. 
Below, we use a simple example to illustrate that locality properties of output points can be used to reduce the number of workers used in coded computation. In general, identifying such properties is highly dependent on the on the function being computed and outside the scope of this work.

Consider the function class $\Fc' = \{h(g(\cdot)), h \in \mathcal{H}, g \in \mathcal{G}\}$ where $\mathcal{G}$ is a set of \textit{non-injective} multivariate polynomials of total degree $d_{\mathcal{G}}$ and $\mathcal{H}$ is a set of multivariate polynomials of total degree $d_{\mathcal{H}}.$ Let $d = d_{\mathcal{G}}d_{\mathcal{H}}$. One can exploit two input points whose images are equal under $g(\cdot)$ to use fewer queries in a coded computation scheme. Suppose that the input points $\langle X_1,\ldots,X_\K\rangle$ are \textit{linearly independent}, the degree $\K-2$ curve $p(z)$ passes through the first $(\K-1)$ input points, and there exists $z_*$ such that $g(p(z_*)) = g(X_\K)$ for all $g \in \mathcal{G}$. Then a scheme which computes $h(g(p(z)))$ for $h \in \mathcal{H}$ and $g \in \mathcal{G}$ by querying workers with evaluation points of $h(g(p(\cdot)))$ and using polynomial interpolation can also decode $h(g(p(z_*))) = h(g(X_\K))$. This reduces the number of workers by at least $d$ compared to the approach of considering a degree $\K-1$ polynomial $p'(z)$ passing through all $\K$ input points and computing $h(g(p'(z)))$.

\subsection{Coded Computation with Byzantine Workers}

\label{sec:adversarialLocal}

As in Section~\ref{sec:corrupt}, we consider a ``coded computation with corruptions'' setting incorporating at most $\byz$ byzantine in addition to $s$ straggling workers. It is simple to modify all proposed coded computation schemes to apply the new domain.

The proposed coded computation schemes are modified as follows: (1) Query each curve at an additional $2\byz$ distinct locations and use the Berlekamp-Welsh algorithm rather than polynomial interpolation \cite{welch1986error}. Choose the query points for the \cics to all be different from the intersection point. (2) Replace the combination of $(s+1)-$wise replication and direct decoding with $(s+2\byz+1)-$wise replication and majority-based decoding. Note that the field size requirements for each scheme increases by at most $2\byz$ in each setting.

\begin{theorem}
Consider the coded computation with corruptions setting in which there are at most $\byz$ byzantine workers and $s$ straggling workers. The modified versions of the \hs and \ns provide robustness to the unreliable workers using $2\byz$ more workers compared to the original schemes, while the \cics does so using $4\byz$ extra workers. Replacing $(s+1)-$wise replication and direct decoding with $(s+2\byz+1)-$wise replication and majority-based decoding likewise ensures resiliency to the unreliable workers and requires an additional $2\byz$ workers per input point.
\end{theorem}

It is simple to extend the modified \hs and \ns to apply to $\K>m$ and $\K>m+1$ input points respectively over $\F^m$. Doing so follows the nearly identical steps as were used in Sections~\ref{sec:homogLoc} and~\ref{sec:nonhomogPoly} to extend the \hs and \ns to apply to $\K>m$ and $\K>m+1$ points; the only exception is in replacing the \hs, \ns, and $(s+1)-$wise replication scheme with the above modified schemes. When the total degree of the polynomial is $(s+2\byz+1)$, number of workers used in this setting by the LCC (i.e. the existing scheme requiring the fewest number of workers) is $\K(s+2\byz+1)$. In contrast, fewer workers are needed by the modified \hs and \ns, as is discussed below.

\begin{remark}
Let function classes $\Fc$ and $\Fc'$ respectively be the sets of homogeneous and non-homogeneous multivariate polynomials of total degree $s+2\byz+1$ over domain $\F^m$. For $\K$ and $\K'$ input points where $(m+1)|\K$ and $(m+2)|\K'$, up to $s$ stragglers, and at most $\byz$ byzantine workers, the \wt under the locality-based model for coded computation is less than or equal to $\K\frac{m}{m+1}(s+2\byz+1)$ and $\K'\frac{m+1}{m+2}(s+2\byz+1)$ respectively.
\end{remark}

\section{Locality-Based Interpretation of Matrix Multiplication Schemes}

So far, we have considered a model of computation in which the objective is to evaluate a function $f$ over \textit{multiple} input points $\langle X_1,\ldots,X_\K\rangle$. Coded computation can also be used for matrix multiplication of a \textit{single} pair of large matrices \cite{lee2017high,yu2017polynomial,dutta2019optimal,baharav2018straggler,kiani2018exploitation,jeong2018locally,yu2018straggler,yu2020entangled}. 
For simplicity of exposition, consider multiplying two square $(\matpartsize \mls \times \matpartsize \mls)$ matrices $A$ and $B$ with entries from a finite field $\F_q$, for some positive integers $\matpartsize$ and $\mls$. In many applications, the overall matrix product is too large for a single worker to compute, but the matrix product subdivides into small modular pieces which can be distributed over several workers. 
One key distinction of this setting is that the process for dividing the computation into smaller components is not prescribed; indeed, the choice of what function(s) the workers compute can impose useful properties for the computation. Nonetheless, existing coded computation schemes for matrix multiplication can be viewed under the locality-based model proposed in this work. 

We now consider two examples of schemes which compute the product of two matrices: the ``polynomial code'' \cite{yu2017polynomial} and the ``MatDot'' code \cite{dutta2019optimal}. Both schemes subdivide the input matrices by rows and columns and then use workers to compute the matrix-matrix products of the (smaller) coded matrices. 
The subdivision of the matrices combined with the matrix-matrix products computed by the workers induce computational locality properties which the schemes then exploit to perform the coded computation.

First, the polynomial code subdivides the computation into the matrix product of the rows of $A$ with the columns of $B$. Thus, $AB$ equals 

$$\begin{bmatrix}
A_1\\
\vdots \\
A_{\matpartsize} \\
\end{bmatrix}
\begin{bmatrix}
B_1 & \cdots & B_\matpartsize\\
\end{bmatrix}
=
\begin{bmatrix}
A_1B_1 & \cdots & A_1 B_\matpartsize \\
\vdots & \ddots & \vdots \\
A_\matpartsize B_1 & \cdots & A_\matpartsize B_\matpartsize\\
\end{bmatrix}.$$

\noindent The workers each compute $f_1(\widetilde{A},\widetilde{B}) = \widetilde{A}\widetilde{B}$ for $(\mls \times \matpartsize \mls)$ matrix $\widetilde{A}$ and $(\matpartsize \mls \times \mls)$ matrix $\widetilde{B}$. At a high level, the scheme computes each $A_iB_j$ for $1 \le i,j \le \matpartsize$, thereby determining $AB$. To do so while tolerating up to $s$ stragglers, the scheme considers the polynomials $p^A_1(z) = \sum_{i=1}^\matpartsize A_i z^{i-1},$ $p^B_1(z) = \sum_{j=1}^\matpartsize B_j z^{(j-1)\matpartsize},$ and $p_1(z) = (p^A_1(z),p^B_1(z))$. The univariate polynomial $f_1(p_1(z)) = \sum_{i=1}^\matpartsize \sum_{j=1}^\matpartsize A_iB_j z^{i-1+(j-1)\matpartsize}$ has each of $A_iB_j$ for $1 \le i,j \le \matpartsize$ as the coefficient for a unique power of $z$. Thus, computing $f_1(p_1(z))$ is sufficient to compute $AB$. The polynomial code computes $f_1(p_1(z))$ using polynomial interpolation via $\matpartsize^2+s$ workers each evaluating $f_1(p_1(z))$ at a unique evaluation point.

Second, the MatDot code divides the computation into the matrix product of the columns of $A$ with the rows of $B$. Thus, $AB$ equals 
$$
\begin{bmatrix}
A_1 & \cdots & A_\matpartsize\\
\end{bmatrix}
\begin{bmatrix}
B_1\\
\vdots \\
B_{\matpartsize} \\
\end{bmatrix}
= \sum_{i=1}^{\matpartsize} A_i B_i.$$
\noindent The workers each compute $f_2(\widetilde{A},\widetilde{B}) = \widetilde{A}\widetilde{B}$ for $(\matpartsize \mls \times \mls)$ matrix $\widetilde{A}$ and $(\mls \times \matpartsize \mls)$ matrix $\widetilde{B}$. At a high level, the scheme computes the polynomial $\left(\sum_{i=1}^{\matpartsize} A_iz^{i-1} \right)\left(\sum_{j=1}^{\matpartsize} B_jz^{\matpartsize-j} \right)$ and decodes the coefficient of $z^{\matpartsize-1}$ as $AB = \sum_{i=1}^{\matpartsize} A_i B_i$. Specifically, the scheme considers the polynomials $p^A_2(z) = \sum_{i=1}^{\matpartsize} A_iz^{i-1}$, $p^B_2(z) = \sum_{j=1}^{\matpartsize} B_jz^{\matpartsize-j},$ and $p_2(z) = (p^A_2(z),p^B_2(z))$. It then interpolates $f_2(p_2(z))$ while tolerating up to $s$ stragglers by using $deg(f_2(p_2(z)) = 2 \matpartsize -1+s$ workers each evaluating $f_2(p_2(z))$ at a unique point. Finally, $AB$ is decoded as the coefficient of $z^{\matpartsize-1}$ of $f_2(p_2(z)).$ 

Under the locality-based model, the polynomial and MatDot codes restricts their attention to the function classes $\Fc_1$ and $\Fc_2$ of multivariate polynomial functions corresponding to matrix-matrix multiplication of  $(\mls \times \matpartsize \mls)$ and $(\matpartsize \mls \times \mls)$ matrices and respectively $(\matpartsize \mls \times \mls)$ and $(\mls \times \matpartsize \mls)$ matrices. Recall the notation from Definition~\ref{def:assCode} that $C^{\Fc_1}$ and $C^{\Fc_2}$ are the associated codes to these function classes; each function in $\Fc_1$ (respectively $\Fc_2$) corresponds to a codeword in $C^{\Fc_1}$ (respectively $C^{\Fc_2}$) which is a vector of all evaluations of the function. 
Consider the curves $P_1 = \{(p^A_1(z),p^B_1(z)) \mid z \in \F_q\}$ and $P_2 = \{(p^A_2(z),p^B_2(z)) \mid z \in \F_q\}.$ For any $\K$ points of $P_1$, there is a is a size $\K$ set $I_1$ such that the punctured code $C^{\Fc_1}_{I_1}$ corresponds to the $\K$ points of $P_1$. Similarly, for any $\K$ points of $P_2$, there is such a size $\K$ corresponding set $I_2$ and punctured code $C^{\Fc_2}{I_2}$. The code symbols $C^{\Fc_1}_{I_1}$ have computational locality $L_{I_1,s} \le \matpartsize^2+s$. This follows from the fact that for all $f_1 \in \Fc_1$, $deg(f_1(p_1(z))=\matpartsize^2-1$\textemdash which is strictly less than $deg(f_1)deg(p_1(z)) = 2\matpartsize(\matpartsize -1)$. The polynomial code is designed to exploit this property. Specifically, it can be viewed as computing the function $f_1$ at the input points $\langle p_1(\alpha_i) \mid 1 \le i \le \matpartsize^2\rangle$ using $\matpartsize^2+s$ workers. Moreover, the computational locality of the code symbols $C^{\Fc_2}_{I_2}$ is $L_{I_2,s} \le 2\matpartsize-1+s$. This holds since for all $f_2 \in \Fc_2$, $deg(f_2(p_2(z))=2\matpartsize-2$. The MatDot code leverages this fact to compute $f_2(p_2(z))$ using interpolation with $2\matpartsize-1+s$ workers while tolerating up to $s$ stragglers. Finally, note that the different choice of subdivision of the matrices by the polynomial code and the MatDot code (leading to different function classes $\Fc_1$ and $\Fc_2$) results in different communication cost in addition to different computational locality properties. 
The function class $\Fc_2$ has a smaller computational locality than that of $\Fc_1$ at a higher communication cost.

In both of the above discussed works~\cite{yu2017polynomial,dutta2019optimal}, the setting of matrix-matrix multiplication of a single pair of large matrices $(A,B)$ using workers capable of only computing smaller (less computationally intensive) matrix-matrix products is considered. This setting naturally extends to the regime of computing the matrix products of multiple pairs of matrices $\langle (A_i,B_i) \mid i \in [\K]\rangle$ using workers with similar computational restrictions, as has been studied in \cite{Jia2019cross,yu2020entangled}. Similar to above, the technique used in this domain is to first divide the matrices in a coded manner and than have the workers compute matrix products of the (smaller) coded matrices. This approach can likewise be interpreted under the locality-based model as a statement on the computational locality of the code symbols related to the matrix partitions.
\section{Conclusion}
\label{sec:conc}

In this work, we introduced a locality-based approach to model coded computation. This model enables the design of coded computation schemes which exploit the well-studied domain of \localRecovery of codes to reduce the required number of workers. 
We demonstrated that the proposed locality-based model for coded computation admits a lower \wt for multivariate polynomials than is possible under the existing \inob approaches. 
This establishes that a factor $s$ overhead in the number of servers\textemdash as in existing approaches\textemdash is not a fundamental requirement for robustness to $s$ stragglers for non-linear functions. The results presented, thus, signal the prospect of future resource efficient coded computation schemes for non-linear functions.

\section*{Acknowledgments}
This work was funded in part by National Science Foundation grant CNS-1850483.
The authors also thank Francisco Maturana for his helpful comments in the writing of this paper.

\bibliographystyle{IEEEtran}
\bibliography{citations}

\end{document}